\documentclass[twocolumn,prb,
preprintnumbers,amsmath,amssymb]{revtex4}

\usepackage[dvips]{graphicx}
\usepackage{dcolumn}
\usepackage{bm}

\begin{document}

\preprint{PRB/}

\title{Verification of polarization selection rules and
implementation of selective coherent manipulations of hydrogenic
transitions in n-GaAs}

\author{M. F. Doty}
\author{B. T. King}
\author{M. S. Sherwin}
 \email{sherwin@physics.ucsb.edu}
\affiliation{California Nanosystems Institute and Institute for
Quantum Engineering, Science, and Technology,\\ University of
California, Santa Barbara.  Santa Barbara, California 93106}
\author{C. R. Stanley}
\affiliation{Department of Electronics and Electrical Engineering,\\
University of Glasgow, Glasgow, G12 8QQ, UK}

\date{\today}

\begin{abstract}
Electrons bound to shallow donors in GaAs have orbital energy levels
analogous to those of the hydrogen atom. The polarization selection
rules for optical transitions between the states analogous to the 1s
and 2p states of hydrogen in a magnetic field are verified using
Terahertz (THz) radiation from the UCSB Free Electron Laser. A
polarization-selective coherent manipulation of the quantum states
is demonstrated and the relevance to quantum information processing
schemes is discussed.
\end{abstract}
\maketitle

In recent years there has been great interest in low dimensional
semiconductors, motivated in large part by the potential to
implement quantum information
processing\cite{Bennett,Gershenfeld,Cirac} in a solid-state
system\cite{Imamoglu, Biolatti, Sherwin}. The development of methods
for selectively addressing and coherently manipulating such quantum
systems is essential to a successful implementation of any quantum
information scheme. In this work we describe verification of the
selection rules for orbital transitions of electrons bound to
shallow donors in GaAs and implementation of selective coherent
manipulations.

The energy levels of shallow donor-bound electrons can be described
by the Coulomb interaction between the electron and the positive
charge at the donor site, modified by the effective mass of the
electron (0.0665 m$_{e}$) and the dielectric constant of the
material (12.56)\cite{Kohn, Klaassen}. This leads to electron energy
levels analogous to those of the hydrogen atom, with a small
correction for the actual donor species, called the central cell
correction (0.110 meV for S donor)\cite{Heron}. The orbital
transitions are in the TeraHertz (THz) frequency regime (a few meV).
In the presence of a magnetic field, additional free electron states
arise, forming a continuum above each Landau level\cite{Klaassen}.
The hydrogenic states can be labelled with the usual notation: $1s$,
$2p^+$, $2p^-$, etc.

The energy of the hydrogenic states can be tuned with an applied
magnetic field to bring orbital transitions into resonance with
fixed frequencies of radiation. Klaassen, et al.\cite{Klaassen} and
Stillman, et al.\cite{Stillman}, have used this technique to perform
spectroscopy experiments. To aid in identifying excited states, they
have assumed that transitions are governed by the usual hydrogen
atom dipole selection rules. However, these experiments have not
included control of the polarization state of the THz beam, and to
our knowledge no work has yet verified that the hydrogen selection
rules remain valid for hydrogenic donors in a semiconductor.

The hydrogen selection rules require a $\sigma^+$-polarized photon
to conserve angular momentum and excite the $1s$ to $2p^+$
transition; $\sigma^-$-polarization is required for $1s$ to $2p^-$.
Since the electron effective mass in GaAs is isotropic, the
selection rules for the bare hydrogen atom might be expected to
hold. However, symmetry in a semiconductor can easily be lost. In
quantum dots, for example, elongation of the dot along specific
crystal planes can break the dot symmetry and mix the polarization
eigenstates of excitons\cite{Stievater}. In this work, we first
verify the selection rules for the $1s$ to $2p^\pm$ transitions by
varying the polarization of the incident THz radiation. We then
utilize the rules to demonstrate a selective coherent manipulation
of orbital energy states.

The sample studied was grown by Molecular Beam Epitaxy (MBE) with a
15-$\mu$m layer of unintentionally-doped GaAs on top of a
semi-insulating GaAs substrate.  The donor density is $2.8 \ \times
\ 10^{14}$ cm$^{-3}$, determined by measuring the Hall resistance,
and sulfur is the dominant impurity. A 200 nm silicon-doped capping
layer was grown on top of the sample to facilitate Ohmic contacts,
and was etched away from the active area of the sample ($50\ \times\
50\ \mu$m$^2$). All experiments were done with the sample immersed
in liquid helium at 1.5 K in a magnetooptical cryostat. At this
temperature, the electrons are initially in the ground state (1s).
Electrons in bound excited states, which have an increased
probability of thermal excitation to the conduction band, are
detected by applying a 50 mV bias and measuring the transient
photoconductivity. The recapture of conduction band electrons occurs
much more rapidly than the repetition rate of our laser ($\sim$1
Hz), and so the electrons return to the ground state before any
subsequent excitation.

\begin{figure}
\resizebox{7.5cm}{!}{\includegraphics{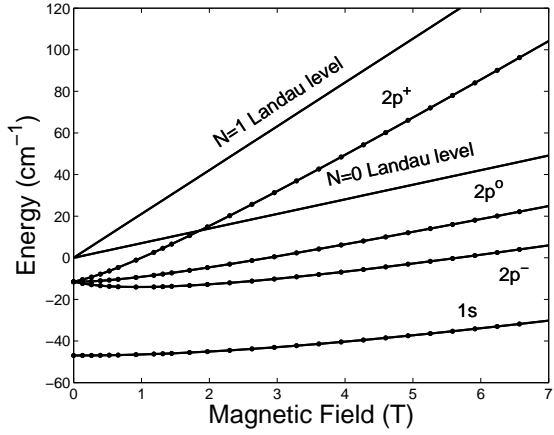}}
\caption{\label{energylevels} {The energy levels of the 1s and 2p
hydrogenic states and the lowest two Landau levels as a function
of magnetic field. The lowest Landau level at zero magnetic field
has been taken to be zero energy. The points are taken from the
calculations of Makado and McGill\cite{MakadoMcGill} and are
converted from dimensionless units using constants given by
Klaassen, et al.\cite{Klaassen} The central cell correction has
been added following the work of Heron, et al. \cite{Heron} The
lines are drawn through the points.}}
\end{figure}

The UCSB Free Electron Laser (FEL) provides linearly polarized THz
radiation for our experiments. Three different THz frequencies in
bands of low atmospheric absorption are used and the hydrogenic
states are tuned into resonance with the radiation using the
magnetic field. Fig. \ref{energylevels} shows the energy of the 1s
and 2p states as a function of magnetic field, as well as the lowest
Landau levels.

Before illuminating the sample, the THz radiation is reflected off a
wire-grid polarizer (rotated by 45 degrees with respect to the
incident electric field) followed by a mirror. The total reflected
beam consists of two orthogonal, linearly-polarized components of
equal amplitude, with a phase delay controlled by the
mirror-polarizer spacing. Varying this spacing varies the
polarization of the light incident on the sample. The metal contacts
on the sample (identical to that used by Cole, et al. \cite{Cole})
are designed to minimize distortions of the incident polarization.

The output polarization state of the light is quantified using the
Stokes parameters: $S_0$-$S_3$. $S_0$, $S_1$ and $S_2$ can be
determined from simple measurements of transmission through an
analyzer as measured with a polarization-insensitive detector, in
our case an InSb hot-electron bolometer. The magnitude of $S_3$
can be determined from the relation
\begin{equation}
S_0^2 = S_1^2 + S_2^2 + S_3^2 \nonumber
\end{equation}
and the sign can be determined by knowing the spacing of the
mirror-polarizer combination with respect to a fixed direction of
linear polarization. The ratio $S_3 / S_0$ is +1 for perfectly
$\sigma^+$-polarized light and -1 for perfectly
$\sigma^-$-polarized light.

\begin{figure}
\resizebox{8.5cm}{!}{\includegraphics{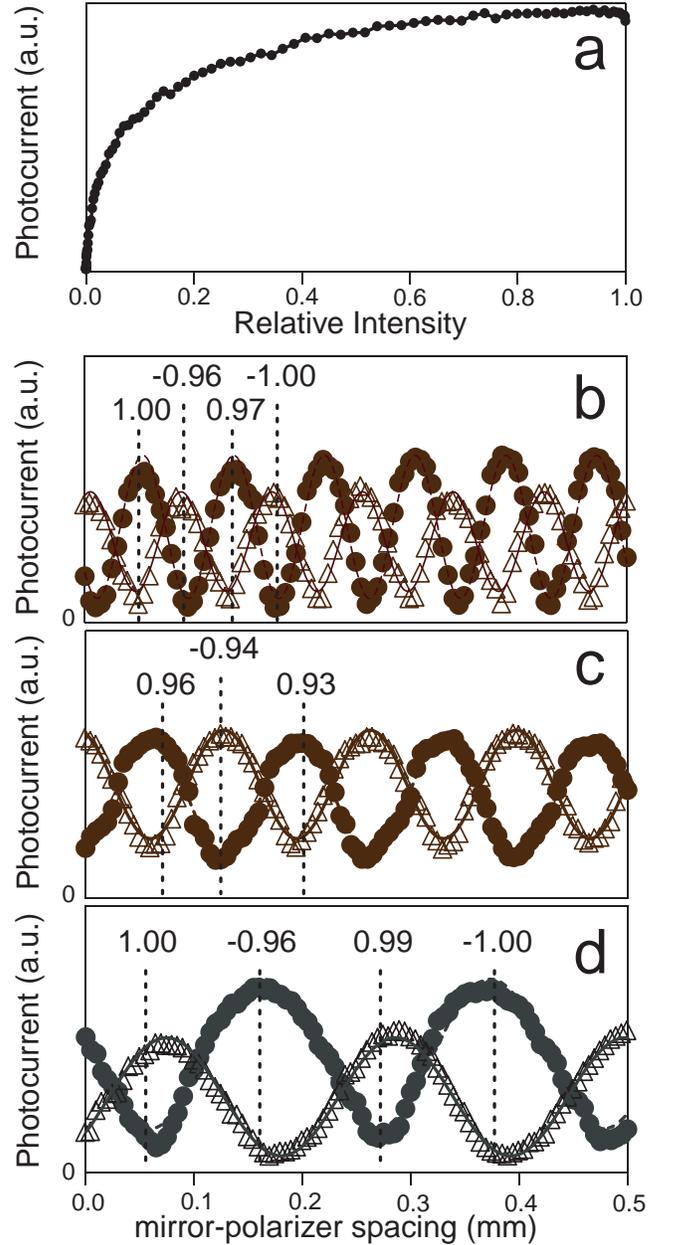}}
\caption{\label{sat_and_selection} {(a) Photocurrent versus relative
intensity (varied with crossed polarizers) of a 5-$\mu$s pulse of
radiation at 1.07 THz. (b-d) Photocurrent as a function of
micrometer spacing (proportional to phase delay). Circles correspond
to a field oriented in the +z direction, triangles to -z. The size
of the symbols is chosen to represent the error bars. The solid and
dashed lines are fits to the data using a sine function. The
vertical cut lines indicate the polarization state of the light
(S$_3/$S$_0$) measured at that spacing. (b) The $1s$ to $2p^+$
transition at 2.53 THz and B = 3.63 T. (c) The $1s$ to $2p^+$
transition at 1.57 THz and B = 1.38 T. (d)  The $1s$ to $2p^-$
transition at 1.01 THz and B = 3.5 T.}}
\end{figure}

To verify the selection rules, the sample is illuminated with a
long ($\sim$5 $\mu$s) pulse of THz radiation and the
photoconductivity is plotted as a function of mirror-polarizer
spacing. The photoconductivity under long-pulse illumination of
this sample always shows a saturation-like behavior as a function
of intensity. See Fig. \ref{sat_and_selection}a for an example.
All data on selection rules is taken with sufficient attenuation
to be in the low-intensity quasi-linear regime.

Fig. \ref{sat_and_selection}b-d shows the photocurrent as a function
of the mirror-polarizer spacing. As expected, the photocurrent
varies with the spacing (polarization of the incident light). The
period of the variation in photocurrent is determined by the change
in mirror-polarizer spacing that corresponds to a one-wavelength
path-length difference. To verify that the change in photocurrent is
due only to the selection rules, the experiment is repeated with the
magnetic field oriented in the +z (circles) and -z (triangles)
direction.

For Fig. \ref{sat_and_selection}b and \ref{sat_and_selection}c, the
maximum photocurrent obtained with the magnetic field in the +z
direction (circles) corresponds to $\sigma^+$-polarized light, the
expected handedness for the 1s to 2p$^+$ transition. For Fig.
\ref{sat_and_selection}d, exciting the 2p$^-$ transition, the
maximum photocurrent corresponds to $\sigma^-$-polarized light. For
all three cases, it is clear that the handedness of polarization
that excites the transition reverses when the static magnetic field
is reversed (triangles), in agreement with the hydrogen atom
selection rules. The sine curves fit to the data for the
anti-parallel directions of the magnetic field should have phases
that differ by $\pi$. The measured phase differences in radians are
(b) 3.63, (c) 3.13 and (d) 3.57. Slight distortion of the incident
polarization (from the focusing element and cryostat windows) is the
most likely source of the offset from zero photoconductivity.
Near-resonant excitation of other transitions may also be
contributing. The difference in maximum signal for the two opposed
directions of the magnetic field is not understood.

The selection rules are now used to perform a selective coherent
state transition. By ``selective transition" we mean that a
coherent manipulation (Rabi oscillation) of the two-level quantum
system is turned on or off by varying only the polarization state
of the incident light. To drive a Rabi oscillation, short, intense
pulses of THz radiation with a controllable pulse width are
generated. The short pulses are extracted from the high-intensity
long-pulse output of the UCSB FEL with laser-activated
semiconductor switches.\cite{Hegmann-1, Hegmann-2} The switches
are manufactured from silicon and are activated by illumination
with near-infrared radiation across the band gap, which generates
a reflecting electron-hole plasma.\cite{Doty} Step-function-like
activation of the switches is obtained with a very short
($\sim$150 fs), intense ($>$1 mJ per switch) near-infrared pulse.
The switches are arranged in a reflection (the reflected light is
kept in the optical path) and transmission (the reflected light is
dumped) geometry. By varying the arrival of the activation pulse
at the reflection and transmission switches, the width of the THz
pulse is controlled.

No attenuation is used for the short pulse experiment since
coherent manipulations require strong coupling. A ``stroboscopic"
measurement, in which the photocurrent is plotted as a function of
the pulse width, is performed to get a map of the electron state
population as a function of time. This method was used by Cole, et
al.\cite{Cole}, who first demonstrated a coherent manipulation of
the orbital states of hydrogenic donors.

\begin{figure}
\centering\resizebox{!}{9cm}{\includegraphics{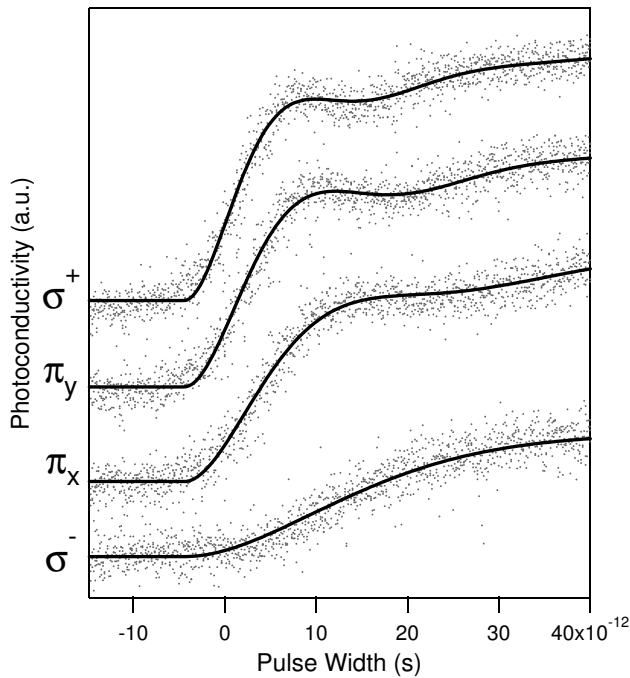}}
\caption{\label{PolarizationResults} {Photoconductivity versus pulse
width using different polarizations. The illumination frequency is
84.5 cm$^{-1}$ (2.54 THz) and the magnetic field is 3.62 T. The
solid lines are fits to the data using the model described in the
text.}}
\end{figure}

Data showing a selective coherent manipulation between the 1s and
2p$^+$ states is presented in Fig. \ref{PolarizationResults}. A
vertical offset has been added to separate the traces. The upper
trace was obtained using $\sigma^+$-polarized light; at pulse widths
of about 15 ps excited electrons are being driven back into the
ground (1s) state before they can ionize and so a decrease in the
photocurrent is observed, the signature of a Rabi oscillation. For
horizontally-polarized ($\pi_x$) and vertically-polarized ($\pi_y$)
light, a Rabi oscillation is still visible, but the period of the
oscillation is longer (20-25 ps) since only half of the
linearly-polarized light couples in to the transition. Small errors
in the mirror-polarizer spacing probably caused the difference in
oscillation period between the horizontally- and
vertically-polarized data. When the polarization is changed to
$\sigma\ ^-$ there is no evidence of a Rabi oscillation. There is a
shallow increase in photoconductivity, which we attribute to a
non-zero $\sigma^+$ component of the polarization arising from the
coupling in to the cryostat, as described above.

Information about the dynamics of the electrons can be extracted
from this data by fitting to a theoretical model. The
density-matrix equations of motion for a two-level system are
used, with damping processes added\cite{Boyd}. The equations are
\begin{eqnarray}
\label{twoleveldensitymatrix}
  \dot{\rho_{11}} &=& \frac{1}{2} i \Omega_R \sigma_{12} - \frac{1}{2} i \Omega_R \sigma_{21} +\Gamma_1\rho_{22} \nonumber \\
  \dot{\rho_{22}} &=& -\frac{1}{2} i \Omega_R \sigma_{12} +\frac{1}{2} i \Omega_R \sigma_{21}- \Gamma_1\rho_{22} - \gamma_3\rho_{22} \nonumber \\
  \dot{\sigma_{12}} &=& -i (\omega - \omega_0) \sigma_{12} + \frac{1}{2} i \Omega_R (\rho_{11} - \rho_{22})- \gamma_2\sigma_{12} \nonumber
\end{eqnarray}
where $\rho_{11}$ and $\rho_{22}$ are proportional to the
populations of the 1s and 2p$^+$ states, respectively.
$\sigma_{12}$ and $\sigma_{21}$($=\sigma_{12}^\ast$) are related
to the phase coherence of the ensemble and are obtained by finding
slowly-varying solutions in the rotating wave approximation.
$\Omega_R = e E_{THz} x_{12} /\hbar$ is the Rabi frequency, with e
the electric charge, $E_{THz}$ the THz electric field amplitude,
and $x_{12}$ the dipole matrix element. The damping terms are:
$\Gamma_1$, the rate of recombination from the 2p$^+$ to the 1s
state (inverse lifetime); $\gamma_2$, the dephasing rate of the
ensemble and $\gamma_3$, the rate of ionization from the 2p$^+$
state into the conduction band. $\omega$ is the frequency of the
driving THz field and $\omega_0$ is the resonant frequency for the
transition.

This model is fit to the data under the assumption that the
driving field is on resonance ($\omega-\omega_0 = 0$) and the
recombination rate is zero ($\Gamma_1=0$). The second assumption
is known to be valid since the recombination rate is much slower
than the other damping processes. The photoconductivity is taken
to be equal to $1-\rho_{11}(t)$ since the 2p$^+$ state lies within
the conduction band and all electrons excited out of the ground
state are expected to arrive in the conduction band within our
probing window. The solid lines in Fig. \ref{PolarizationResults}
show the results of the fits. The fit to the $\sigma^+$ data
yields the values $\gamma_2 = 1.98\ \times\ 10^{11}\ s^{-1}$ and
$\gamma_3 = 1.22\ \times\ 10^{11}\ s^{-1}$. These values
correspond to a dephasing time of 5 ps and ionization time of 8
ps. Considerably longer dephasing and ionization times were
reported by Cole, et al.\cite{Cole} for illumination with lower
intensities of THz radiation. The longer damping time arises
because the dominant damping process is associated with coupling
to the conduction band states, as described by Brandi, et
al.\cite{Brandi}. Lower intensities result in slower excitation to
the conduction band and thus slower contribution of this extrinsic
damping.

In summary, the selection rules for hydrogenic transitions in n-GaAs
have been verified by monitoring the photocurrent, which is
generated by ionization of the bound excited state population, as a
function of incident polarization. These selection rules can be used
to determine the required electric fields in the design of a THz
cavity to couple hydrogenic states of different donors. A THz
photonic cavity, which is significantly larger than an equivalent
visible-frequency cavity, would facilitate spatial addressing of
individual donors in a quantum information processing
scheme\cite{Sherwin}.

Using short, intense pulses of THz radiation polarized in compliance
with the selection rules, Rabi oscillations were driven between
orbital energy states. The absence of a Rabi oscillation when
illuminating with high-intensity but incorrectly-polarized radiation
demonstrates a selective state manipulation. Polarized radiation
that coupled specific states was used by Li, et al. to demonstrate a
quantum logic gate using excitons in a single quantum dot.\cite{Li}
The selective coherent manipulation presented in this work is a
first step toward the implementation of more complex quantum logic
operations using donor-bound electrons.

\begin{acknowledgments}
The authors wish to acknowledge the assistance of David Enyeart in
the operation of the UCSB FEL. This work was supported by the DARPA
QuIST program under Grant No. MDA972-01-1-0027 and by SUN
microsystems.
\end{acknowledgments}


\begin{thebibliography}{19}
\expandafter\ifx\csname
natexlab\endcsname\relax\def\natexlab#1{#1}\fi
\expandafter\ifx\csname bibnamefont\endcsname\relax
  \def\bibnamefont#1{#1}\fi
\expandafter\ifx\csname bibfnamefont\endcsname\relax
  \def\bibfnamefont#1{#1}\fi
\expandafter\ifx\csname citenamefont\endcsname\relax
  \def\citenamefont#1{#1}\fi
\expandafter\ifx\csname url\endcsname\relax
  \def\url#1{\texttt{#1}}\fi
\expandafter\ifx\csname
urlprefix\endcsname\relax\def\urlprefix{URL }\fi
\providecommand{\bibinfo}[2]{#2}
\providecommand{\eprint}[2][]{\url{#2}}

\bibitem[{\citenamefont{{C. H. Bennett} and {D. P.
  DiVincenzo}}(2000)}]{Bennett}
\bibinfo{author}{\bibnamefont{{C. H. Bennett}}} \bibnamefont{and}
  \bibinfo{author}{\bibnamefont{{D. P. DiVincenzo}}}, \bibinfo{journal}{Nature}
  \textbf{\bibinfo{volume}{404}}, \bibinfo{pages}{247} (\bibinfo{year}{2000}).

\bibitem[{\citenamefont{{N. A. Gershenfeld} and {I. L.
  Chuang}}(1997)}]{Gershenfeld}
\bibinfo{author}{\bibnamefont{{N. A. Gershenfeld}}} \bibnamefont{and}
  \bibinfo{author}{\bibnamefont{{I. L. Chuang}}}, \bibinfo{journal}{Science}
  \textbf{\bibinfo{volume}{275}}, \bibinfo{pages}{350} (\bibinfo{year}{1997}).

\bibitem[{\citenamefont{{J. I. Cirac} and {P. Zoller}}(1995)}]{Cirac}
\bibinfo{author}{\bibnamefont{{J. I. Cirac}}} \bibnamefont{and}
  \bibinfo{author}{\bibnamefont{{P. Zoller}}}, \bibinfo{journal}{Phys. Rev.
  Lett} \textbf{\bibinfo{volume}{74}}, \bibinfo{pages}{4091}
  (\bibinfo{year}{1995}).

\bibitem[{\citenamefont{{A. Imamoglu} et~al.}(1999)\citenamefont{{A. Imamoglu},
  {D. D. Awschalom}, {G. Burkard}, {D. P. DiVincenzo}, {D. Loss}, {M. Sherwin},
  and {A. Small}}}]{Imamoglu}
\bibinfo{author}{\bibnamefont{{A. Imamoglu}}},
  \bibinfo{author}{\bibnamefont{{D. D. Awschalom}}},
  \bibinfo{author}{\bibnamefont{{G. Burkard}}},
  \bibinfo{author}{\bibnamefont{{D. P. DiVincenzo}}},
  \bibinfo{author}{\bibnamefont{{D. Loss}}}, \bibinfo{author}{\bibnamefont{{M.
  Sherwin}}}, \bibnamefont{and} \bibinfo{author}{\bibnamefont{{A. Small}}},
  \bibinfo{journal}{Phys. Rev. Lett.} \textbf{\bibinfo{volume}{83}},
  \bibinfo{pages}{4204} (\bibinfo{year}{1999}).

\bibitem[{\citenamefont{{E. Biolatti} et~al.}(2000)\citenamefont{{E. Biolatti},
  {R. C. Iotti}, {P. Zanardi}, and {F. Rossi}}}]{Biolatti}
\bibinfo{author}{\bibnamefont{{E. Biolatti}}},
  \bibinfo{author}{\bibnamefont{{R. C. Iotti}}},
  \bibinfo{author}{\bibnamefont{{P. Zanardi}}}, \bibnamefont{and}
  \bibinfo{author}{\bibnamefont{{F. Rossi}}}, \bibinfo{journal}{Phys. Rev.
  Lett.} \textbf{\bibinfo{volume}{85}}, \bibinfo{pages}{5647}
  (\bibinfo{year}{2000}).

\bibitem[{\citenamefont{{M.S. Sherwin} et~al.}(1999)\citenamefont{{M.S.
  Sherwin}, {A. Imamoglu}, and {T. Montroy}}}]{Sherwin}
\bibinfo{author}{\bibnamefont{{M.S. Sherwin}}},
  \bibinfo{author}{\bibnamefont{{A. Imamoglu}}}, \bibnamefont{and}
  \bibinfo{author}{\bibnamefont{{T. Montroy}}}, \bibinfo{journal}{Phys. Rev. A}
  \textbf{\bibinfo{volume}{60}}, \bibinfo{pages}{3508} (\bibinfo{year}{1999}).

\bibitem[{\citenamefont{Kohn}(1957)}]{Kohn}
\bibinfo{author}{\bibfnamefont{W.}~\bibnamefont{Kohn}}, \bibinfo{journal}{Solid
  State Physics: Advances in Research and Applications}
  \textbf{\bibinfo{volume}{5}}, \bibinfo{pages}{257} (\bibinfo{year}{1957}).

\bibitem[{\citenamefont{{Tjeerd O. Klaassen} et~al.}(1998)\citenamefont{{Tjeerd
  O. Klaassen}, {Janette L. Dunn}, and {Colin A. Bates}}}]{Klaassen}
\bibinfo{author}{\bibnamefont{{Tjeerd O. Klaassen}}},
  \bibinfo{author}{\bibnamefont{{Janette L. Dunn}}}, \bibnamefont{and}
  \bibinfo{author}{\bibnamefont{{Colin A. Bates}}}, in
  \emph{\bibinfo{booktitle}{Atoms and Molecules in Strong External Fields}},
  edited by \bibinfo{editor}{\bibnamefont{Schmelcher}} \bibnamefont{and}
  \bibinfo{editor}{\bibnamefont{Schweizer}} (\bibinfo{publisher}{Plenum Press},
  \bibinfo{year}{1998}), pp. \bibinfo{pages}{291--300}.

\bibitem[{\citenamefont{{R.J. Heron} et~al.}(1999)\citenamefont{{R.J. Heron},
  {R.A. Lewis}, {P.E. Simmonds}, {R.P. Starrett}, {A.V. Skougarevsky}, {R.G.
  Clark}, and {C.R. Stanley}}}]{Heron}
\bibinfo{author}{\bibnamefont{{R.J. Heron}}},
  \bibinfo{author}{\bibnamefont{{R.A. Lewis}}},
  \bibinfo{author}{\bibnamefont{{P.E. Simmonds}}},
  \bibinfo{author}{\bibnamefont{{R.P. Starrett}}},
  \bibinfo{author}{\bibnamefont{{A.V. Skougarevsky}}},
  \bibinfo{author}{\bibnamefont{{R.G. Clark}}}, \bibnamefont{and}
  \bibinfo{author}{\bibnamefont{{C.R. Stanley}}}, \bibinfo{journal}{Journal of
  Applied Physics} \textbf{\bibinfo{volume}{85}}, \bibinfo{pages}{893}
  (\bibinfo{year}{1999}).

\bibitem[{\citenamefont{{G.E. Stillman} et~al.}(1969)\citenamefont{{G.E.
  Stillman}, {C.M. Wolfe}, and {J.O. Dimmock}}}]{Stillman}
\bibinfo{author}{\bibnamefont{{G.E. Stillman}}},
  \bibinfo{author}{\bibnamefont{{C.M. Wolfe}}}, \bibnamefont{and}
  \bibinfo{author}{\bibnamefont{{J.O. Dimmock}}}, \bibinfo{journal}{Solid State
  Communications} \textbf{\bibinfo{volume}{7}}, \bibinfo{pages}{921}
  (\bibinfo{year}{1969}).

\bibitem[{\citenamefont{{T. H. Stievater} et~al.}(2002)\citenamefont{{T. H.
  Stievater}, {X. Li}, {T .Cubel}, {D. G. Steel}, {D. Gammon}, {D. S. Katzer},
  and {D. Park}}}]{Stievater}
\bibinfo{author}{\bibnamefont{{T. H. Stievater}}},
  \bibinfo{author}{\bibnamefont{{X. Li}}}, \bibinfo{author}{\bibnamefont{{T
  .Cubel}}}, \bibinfo{author}{\bibnamefont{{D. G. Steel}}},
  \bibinfo{author}{\bibnamefont{{D. Gammon}}},
  \bibinfo{author}{\bibnamefont{{D. S. Katzer}}}, \bibnamefont{and}
  \bibinfo{author}{\bibnamefont{{D. Park}}}, \bibinfo{journal}{App. Phys.
  Lett.} \textbf{\bibinfo{volume}{81}}, \bibinfo{pages}{4251}
  (\bibinfo{year}{2002}).

\bibitem[{\citenamefont{{X. Li} et~al.}(2003)\citenamefont{{X. Li}, {Y. Wu},
  {D. Steel}, {D. Gammon}, {T. H. Stievater}, {D. S. Katzer}, {D. Park}, {C.
  Piermarocchi}, and {L. J. Sham}}}]{Li}
\bibinfo{author}{\bibnamefont{{X. Li}}}, \bibinfo{author}{\bibnamefont{{Y.
  Wu}}}, \bibinfo{author}{\bibnamefont{{D. Steel}}},
  \bibinfo{author}{\bibnamefont{{D. Gammon}}},
  \bibinfo{author}{\bibnamefont{{T. H. Stievater}}},
  \bibinfo{author}{\bibnamefont{{D. S. Katzer}}},
  \bibinfo{author}{\bibnamefont{{D. Park}}}, \bibinfo{author}{\bibnamefont{{C.
  Piermarocchi}}}, \bibnamefont{and} \bibinfo{author}{\bibnamefont{{L. J.
  Sham}}}, \bibinfo{journal}{Science} \textbf{\bibinfo{volume}{301}},
  \bibinfo{pages}{809} (\bibinfo{year}{2003}).

\bibitem[{\citenamefont{{P.C. Makado} and {N.C. McGill}}(1986)}]{MakadoMcGill}
\bibinfo{author}{\bibnamefont{{P.C. Makado}}} \bibnamefont{and}
  \bibinfo{author}{\bibnamefont{{N.C. McGill}}}, \bibinfo{journal}{J. Phys. C:
  Solid State Phys.} \textbf{\bibinfo{volume}{19}}, \bibinfo{pages}{873}
  (\bibinfo{year}{1986}).

\bibitem[{\citenamefont{{B.E. Cole} et~al.}(2001)\citenamefont{{B.E. Cole},
  {J.B. Williams}, {B.T. King}, {M.S. Sherwin}, and {C.R. Stanley}}}]{Cole}
\bibinfo{author}{\bibnamefont{{B.E. Cole}}},
  \bibinfo{author}{\bibnamefont{{J.B. Williams}}},
  \bibinfo{author}{\bibnamefont{{B.T. King}}},
  \bibinfo{author}{\bibnamefont{{M.S. Sherwin}}}, \bibnamefont{and}
  \bibinfo{author}{\bibnamefont{{C.R. Stanley}}}, \bibinfo{journal}{Nature}
  \textbf{\bibinfo{volume}{410}}, \bibinfo{pages}{60} (\bibinfo{year}{2001}).

\bibitem[{\citenamefont{Hegmann and Sherwin}(1996)}]{Hegmann-1}
\bibinfo{author}{\bibfnamefont{F.~A.} \bibnamefont{Hegmann}} \bibnamefont{and}
  \bibinfo{author}{\bibfnamefont{M.~S.} \bibnamefont{Sherwin}}, in
  \emph{\bibinfo{booktitle}{International conference on millimeter and
  submillimeter waves and applications III.}} (\bibinfo{organization}{SPIE},
  \bibinfo{year}{1996}).

\bibitem[{\citenamefont{{F. A. Hegmann} et~al.}(2000)\citenamefont{{F. A.
  Hegmann}, {J. B. Williams}, {B. Cole}, {M. S. Sherwin}, {J. W. Beeman}, and
  {E. E. Haller}}}]{Hegmann-2}
\bibinfo{author}{\bibnamefont{{F. A. Hegmann}}},
  \bibinfo{author}{\bibnamefont{{J. B. Williams}}},
  \bibinfo{author}{\bibnamefont{{B. Cole}}}, \bibinfo{author}{\bibnamefont{{M.
  S. Sherwin}}}, \bibinfo{author}{\bibnamefont{{J. W. Beeman}}},
  \bibnamefont{and} \bibinfo{author}{\bibnamefont{{E. E. Haller}}},
  \bibinfo{journal}{Appl. Phys. Lett.} \textbf{\bibinfo{volume}{76}},
  \bibinfo{pages}{262} (\bibinfo{year}{2000}).

\bibitem[{\citenamefont{{M. F. Doty} et~al.}(2004)\citenamefont{{M. F. Doty},
  {B. E. Cole}, {B. T. King}, and {M. S. Sherwin}}}]{Doty}
\bibinfo{author}{\bibnamefont{{M. F. Doty}}}, \bibinfo{author}{\bibnamefont{{B.
  E. Cole}}}, \bibinfo{author}{\bibnamefont{{B. T. King}}}, \bibnamefont{and}
  \bibinfo{author}{\bibnamefont{{M. S. Sherwin}}}, \bibinfo{journal}{Rev. Sci.
  Inst.} \textbf{\bibinfo{volume}{75}}, \bibinfo{pages}{2921}
  (\bibinfo{year}{2004}).

\bibitem[{\citenamefont{Boyd}(1992)}]{Boyd}
\bibinfo{author}{\bibfnamefont{R.~W.} \bibnamefont{Boyd}},
  \emph{\bibinfo{title}{Nonlinear Optics}} (\bibinfo{publisher}{Academic
  Press}, \bibinfo{year}{1992}).

\bibitem[{\citenamefont{{H.S. Brandi} et~al.}(2003)\citenamefont{{H.S. Brandi},
  {A. Latge}, and {L.E. Oliveira}}}]{Brandi}
\bibinfo{author}{\bibnamefont{{H.S. Brandi}}},
  \bibinfo{author}{\bibnamefont{{A. Latge}}}, \bibnamefont{and}
  \bibinfo{author}{\bibnamefont{{L.E. Oliveira}}}, \bibinfo{journal}{Phys. Rev.
  B} \textbf{\bibinfo{volume}{68}}, \bibinfo{pages}{233206}
  (\bibinfo{year}{2003}).

\end{thebibliography}
\end{document}